\documentclass[aps,onecolumn,11pt,showpacs,preprintnumbers]{revtex4-1}

\usepackage{amsmath}
\usepackage{graphicx}
\usepackage{epstopdf}
\usepackage{hyperref}
\usepackage{pgfplots}

\newcommand{\ai}{a_{i}}
\newcommand{\Ui}{\Upsilon_{i}}
\newcommand{\gi}{\gamma_{i}}
\newcommand{\Hi}{H_{i}}
\newcommand{\Ti}{T_{i}}
\newcommand{\si}{s_{i}}
\newcommand{\aRH}{a_{\rm RH}}
\newcommand{\fRH}{\Upsilon_{\rm RH}}

\newcommand{\HRH}{H_{\rm RH}}
\newcommand{\TRH}{T_{\rm RH}}

\newcommand{\Tkd}{T_{\rm kd}}
\newcommand{\Tkdb}{T_{\rm kd}^{\rm BH} }
\newcommand{\Tkdv}{T_{\rm kd}^{\rm VG} }

\newcommand{\Tstd}{T_{\rm kd, std}}
\newcommand{\hs}{\hat{s}}
\newcommand{\hsR}{\hat{s}_R}

\begin{document}

\title{Kinetic decoupling of WIMPs: analytic expressions}

\author{Luca Visinelli}
\email{luca.visinelli@utah.edu}
\author{Paolo Gondolo}
\email{paolo.gondolo@utah.edu}
\affiliation{Department of Physics and Astronomy, University of Utah, 115 South 1400 East \#201, Salt Lake City, Utah 84112-0830, USA}

\date{\today}

\begin{abstract}
\noindent
We present a general expression for the values of the average kinetic energy and of the temperature of kinetic decoupling of a WIMP, valid for any cosmological model. We show an example of the usage of our solution when the Hubble rate has a power-law dependence on temperature, and we show results for the specific cases of kination cosmology and low-temperature reheating cosmology.
\end{abstract}


\maketitle

\section{Introduction}

Despite the various astrophysical observations in support of its existence~\cite{komatsu, planck}, the nature of dark matter still remains an open question. Of the various candidates for dark matter, one of the most compelling is the Weakly Interacting Massive Particle (WIMP)~\cite{zwicky, kolb_book, jungman1996, bertone2005, kuhlen2012}, with a mass ranging from a few GeV to $10{\rm~TeV}$. In fact, when the WIMP annihilation rate falls below the Hubble expansion rate, the chemical equilibrium between WIMPs and the primordial plasma is no longer maintained, and the number of WIMPs per comoving volume naturally fixes to the value required for explaining the present abundance of cold dark matter. Although chemical equilibrium at this stage is no longer maintained, kinetic equilibrium between dark matter and the plasma is still achieved through a high momentum exchange rate~\cite{hofmann, chen, berezinsky, green, bertschinger, bringmann, kasahara, bi, gondolo}. Eventually, when the Hubble rate equates the scattering process rate, WIMPs kinetically decouple from the plasma and flow with a given free-streaming velocity. This velocity sets the lowest value for the size of protohalos, which determines the subsequent evolution of primordial structures~\cite{schmid, boehm, loeb, profumo, gondolo_gelmini, aarssen, cornell}. In particular, Bringmann~\cite{bringmann} defined the temperature of the kinetic decoupling $T_{\rm kd}$ in the standard cosmological scenario, while Gelmini and Gondolo~\cite{gondolo_gelmini} defined $T_{\rm kd}$ in the Low-Temperature Reheating (LTR) cosmology following a dimensionality reasoning.

In this paper, we present a full solution of the evolution equation governing the process of the kinetic decoupling, and we generalize the definition of the temperature of kinetic decoupling and the average kinetic energy of WIMPs in a generic non-standard cosmological model. This paper is organized as follows. In Sec.~\ref{General equation}, we solve the evolution equation for the WIMP kinetic energy in a generic cosmological background, and we apply our results to the standard radiation-dominated scenario in Sec.~\ref{Standard cosmology}, to a power-law cosmology in Sec~\ref{Power-law models}, and to a broken power law cosmology in Sec.~\ref{Broken power-law model}. In particular, we discuss the Low Temperature Reheating (LTR) scenario in Sec.~\ref{LTR} and the kination scenario in Sec.~\ref{kination}. We summarize our results in Sec.~\ref{Summary}.

\section{General solution of the temperature equation for Dark Matter in a thermal bath} \label{General equation}

The scattering process between plasma at temperature $T$ and WIMPs of mass $M_\chi \gg T$ can be described by a Brownian motion in momentum space. The momentum transfer $\Delta p$ and the average momentum of the dark matter particles $p$ are related by
\begin{equation}
p = \sqrt{N_e}\,\Delta p,
\end{equation}
where $N_e$ is the number of collisions required to change the momentum by $p$. Since $p \sim \sqrt{M_\chi\,T}$ is much larger than the average momentum transfer $\Delta p \sim T$, the number of collisions required to appreciably change the momentum of WIMP is $N_e = (p/\Delta p)^2\sim M_\chi/T \gg 1$. The momentum exchange rate $\Gamma$ is suppressed with respect to the elastic collision rate $\Gamma_{\rm el}$ by a factor $T/M_\chi$. Thermal decoupling of WIMPs occurs at a temperature $\Tkd$ approximatively given by $H(\Tkd) \sim \Gamma$, where $H = H(T)$ is the Hubble expansion rate at temperature $T$. A more precise way to define $\Tkd$ consists in using Boltzmann's equation \cite{hofmann, berezinsky, green, bertschinger, bringmann, kasahara, gondolo}, which does not require the dark matter to be treated as a perfect fluid. In particular, Ref.~\cite{kasahara} discussed the Boltzmann equation under the generic assumptions of a heavy dark matter particle with $M_\chi \gg T$ and with small momentum transfer per collision $\Delta p \ll p$, obtaining a Fokker-Planck equation for the dark matter particle occupation number $f_\chi = f_\chi({\bf p}_\chi)$,
\begin{equation} \label{eq:boltzmann}
\frac{\partial f_\chi}{\partial t} - H(T)\,{\bf p}_\chi\,\cdot\,\frac{\partial f_\chi}{\partial {\bf p}_\chi} = \gamma(T)\, \frac{\partial}{\partial {\bf p}_\chi}\,\cdot\,\left({\bf p}_\chi\,f_\chi\,(1\pm f_\chi) + M_\chi\,T\, \frac{\partial f_\chi}{\partial {\bf p}_\chi}\right).
\end{equation}
The momentum relaxation rate $\gamma(T)$ is defined in terms of the scattering cross section between the WIMP and each relativistic species in the plasma at temperature $T$, see Eq.~(2) in Ref.~\cite{gondolo}. Here we are not interested in the specific form of $\gamma(T)$, since we are interested in a general solution to Eq.~(\ref{eq:boltzmann}). We only assume that $\gamma(T)$ monotonically increases with $T$.

Defining the WIMP kinetic temperature $T_\chi$ as 2/3 of the average kinetic energy of the dark matter particle,
\begin{equation}
T_\chi = \frac{2}{3}\,\int \frac{{\bf p}_\chi^2}{2M_\chi}\,f_\chi({\bf p}_\chi)\,d^3{\bf p}_\chi,
\end{equation} 
and using the Fokker-Planck Eq.~(\ref{eq:boltzmann}) with the approximation $1 \pm f_\chi \approx 1$ and the Hubble rate $H = \dot{a}/a$, it can be shown that $T_\chi$ satisfies the differential equation~\cite{kasahara, gondolo}
\begin{equation} \label{mastereq}
a\,\frac{dT_\chi}{da} + 2 T_\chi = -\frac{2\gamma(a)}{H}\,(T_\chi-T).
\end{equation}
Eq.~(\ref{mastereq}) is the central equation we consider in this paper. Refs.~\cite{bertschinger} and~\cite{bringmann} present analytic solutions to Eq.~(\ref{mastereq}) in the standard cosmology and for $\gamma(T)$ proportional to a power of $T$, while in Ref.~\cite{gondolo} Eq.~(\ref{mastereq}) is solved numerically in the standard cosmology and in the presence of quark interactions. Here, we solve Eq.~(\ref{mastereq}) in terms of analytic expressions for a generic cosmological model.

The scattering of the WIMP particles off the plasma is regulated by the function
\begin{equation} \label{def_f}
\Upsilon(T) = \frac{\gamma(T)}{H(T)},
\end{equation}
in terms of which we rewrite Eq.~(\ref{mastereq}) as
\begin{equation} \label{mastereq1}
a\,\frac{dT_\chi}{da} + 2\left[1+\Upsilon(T)\right]\,T_\chi = 2\Upsilon(T)\,T.
\end{equation}
The general solution for $T_\chi$ must satisfy the following boundary conditions. For $\gamma(T) \gg H(T)$, WIMPs are tightly coupled to the plasma and the kinetic temperature approaches 
\begin{align}
a\,T_\chi = {\rm constant},
\end{align}
from which, using Eq.~(\ref{mastereq1}), we obtain the behavior
\begin{equation} \label{lim_gamma}
T - T_\chi \approx \frac{T_\chi}{2\Upsilon(T)} \to 0,\quad \hbox{or}\quad T_\chi \approx T,\quad \hbox{for $T \to \infty$}.
\end{equation}
In the opposite limit $\gamma(T) \ll H(T)$, WIMPs decouple from the plasma, and Eq.~(\ref{mastereq}) reduces to 
\begin{equation} \label{decoupled}
a\,\frac{dT_\chi}{da} + 2 T_\chi = 0,
\end{equation}
with solution 
\begin{align}
a^2\,T_\chi = {\rm const}.
\end{align}

\subsection{Analytic expression for the kinetic temperature $T_\chi$}

We solve Eq.~(\ref{mastereq1}) by the method of undetermined coefficients, by first considering the homogeneous equation associated with it,
\begin{equation} \label{mastereq1_hom}
a\,\frac{d\,T_\chi^{(\rm hom)}}{da} + 2\left[1+\Upsilon(T)\right]\,T_\chi^{(\rm hom)} = 0,
\end{equation}
whose solution from the initial value of the scale factor $a_i$ to $a$ is
\begin{equation}
T^{(\rm hom)}_\chi(a) = T^{(\rm hom)}_i\,\left(\frac{a_i}{a}\right)^2\,e^{-G(a,\ai)}.
\end{equation}
Here, $T^{(\rm hom)}_i = T^{(\rm hom)}_\chi(a_i)$ is a constant temperature that will disappear from the full solution, and we have defined the function
\begin{equation}\label{def_G}
G(a,a') = 2\int_{a'}^a\,\Upsilon(a'')\, \frac{da''}{a''},
\end{equation}
which satisfies the relation
\begin{equation} \label{property_G}
G(a, a') = G(a,a'') + G(a'',a'), \quad\hbox{for $a \leq a'' \leq a'$}.
\end{equation}
Notice that, when expressed as a function of time, Eq.~(\ref{def_G}) is simply given by
\begin{equation}\label{def_G_t}
G(t,t') = 2\int_{t'}^t\,\gamma(t'')\,dt''.
\end{equation}
The particular solution to Eq.~(\ref{mastereq1}) is obtained by using the method of undetermined coefficients,
\begin{equation}
T_\chi^{(\rm part)}(a) = T_\chi^{(\rm hom)}(a)\,\hat{u}(a),
\end{equation}
where the function $\hat{u}(a)$ satisfies
\begin{equation} \label{mastereq2}
a\,T_\chi^{(\rm hom)}(a)\,\frac{d\hat{u}(a)}{da} = 2\Upsilon(a)\,T(a),
\end{equation}
The solution to Eq.~(\ref{mastereq2}) is
\begin{equation}
T_\chi^{(\rm part)}(a) = \frac{2}{a^{2}}\,\int_{a_i}^a\,e^{-G(a, a')}\,\Upsilon(a')\,T(a')\,a'\,da'.
\end{equation}
The complete solution to Eq.~(\ref{mastereq1}) that satisfies the condition $T_\chi(a_i) = T_{\chi i}$ is the sum of the homogeneous and particular solutions,
\begin{equation} \label{sol2}
T_\chi(a) = T_{\chi i}\,\left(\frac{a_i}{a}\right)^2\,e^{-G(a,\ai)} + \frac{2}{a^{2}}\,\int_{a_i}^a\,e^{-G(a, a')}\,\Upsilon(a')\,T(a')\,a'\,da'.
\end{equation}
Eq.~(\ref{sol2}) provides the value of the kinetic temperature $T_\chi$ as a function of the temperature of the Universe $T$ for a generic cosmological model.
Integration by parts of Eq.~(\ref{sol2}) using $de^{-G(a,a')}=e^{-G(a,a')}2\Upsilon(a')da'/a'$ gives the alternative form
\begin{align} \label{sol2_3}
T_\chi(a) = T(a) + \left( T_{\chi i}-T_i \right) \,\left(\frac{a_i}{a}\right)^2\,e^{-G(a,\ai)} - \int_{a_i}^a e^{-G(a,a')}\,\frac{d}{da'}\left[\left(\frac{a'}{a}\right)^2\,T(a')\right]\,da' .
\end{align}
Here $T_i=T(a_i)$ is the initial plasma temperature.

Another alternative form of the solution is obtained by introducing the indefinite integral 
\begin{align} \label{gen_def_s}
s(a) = 2 \int^{a} \frac{\gamma(a')}{H(a')} \, \frac{da'}{a'} ,
\end{align}
which is defined apart from a constant that disappears from the solution. The integral in Eq.~(\ref{def_G}) can then be written as
\begin{equation} \label{eq_G_powerlaw}
G(a,a') = s(a) - s(a').
\end{equation}
Eq.~(\ref{sol2}) in terms of the variable $s = s(a)$ is rewritten as
\begin{equation} \label{sol3}
T_\chi = T_{\chi i}\,\left(\frac{\ai}{a}\right)^2\,e^{s-\si} - \int_{\si}^s\,\left(\frac{a'}{a}\right)^2\,e^{s - s'}\,T(s')\,ds',
\end{equation}
where $\si = s(\ai)$.

When the initial condition $a_i \to 0$, the exponential term in the homogeneous solution of Eq.~(\ref{sol2}) drops to zero, while the product $a_i\,T_i$ remains constant because of the limit in Eq.~(\ref{lim_gamma}), yielding
\begin{equation} \label{sol2_2}
T_\chi(a) = \frac{2}{a^{2}}\,\int_{0}^{a}\,e^{-G(a, a')}\,\Upsilon(a')\,T(a')\,a'\,da' , \qquad \text{if } a_i\to0.
\end{equation}
Equivalently from Eq.~(\ref{sol2_3}), using $T_{\chi i}=T_i$ at very small $a_i$,
\begin{equation} \label{behavior_asymptotic}
T_\chi(a) = T(a) - \int_{0}^{a}\,e^{-G(a,a')}\,\frac{d}{da'}\left[\left(\frac{a'}{a}\right)^2\,T(a')\right]\,da', \qquad \text{if $a_i\to0$,  $T_{\chi i}=T_i$.}
\end{equation}
Since $a'\,T(a') \to {\rm const}$ for $a'\to 0$, the derivative in the integrand remains finite in the limit of tight coupling, while the exponential term drops to zero because of the limit in Eq.~(\ref{lim_gamma}). Thus, $T_\chi \approx T$ in the tightly coupled limit.

\subsection{Temperature of kinetic decoupling} \label{Temperature of kinetic decoupling}

The temperature of kinetic decoupling $\Tkd$ expresses the temperature of the plasma at which the kinetic decoupling of WIMPs occurs. Here, we use the definition~\cite{gondolo},
\begin{equation} \label{relation}
\gamma(\Tkd) = H(\Tkd),
\end{equation}
where $H(\Tkd)$ is the Hubble expansion rate when WIMPs decouple kinetically from the primordial plasma. In the literature, different definitions of the temperature of kinetic decoupling can be found. In Bertschinger~\cite{bertschinger}, the definition of the temperature $\Tkd^B$ differs from our Eq.~\eqref{relation} by a factor of two,
\begin{equation} \label{relation_bertschinger}
\gamma(\Tkd^B) = 2\,H(\Tkd^B),
\end{equation}
while another definition of $\Tkd$ is given by equating the rate of heat transfer equal to the Hubble expansion rate, $2\gamma(\Tkd)=H(\Tkd)$.

Bringmann and Hofmann~\cite{bringmann} define the temperature of kinetic decoupling as
\begin{equation} \label{Tkd_bringmann}
\Tkdb = \frac{T^2}{T_\chi}\bigg|_{T\to 0}.
\end{equation}
Although the four definitions for $\Tkd$ yield the same dependence on $T_\chi$, up to a numerical constant, there are some conceptual differences between them. The definition in Eq.~(\ref{Tkd_bringmann}) depends in principle on the moment at which the temperature of KD is computed, and requires knowledge of the evolution of the universe at late times, far after kinetic decoupling. Instead, the definition we adopted in Eq.~(\ref{relation}) and Eq.~\eqref{relation_bertschinger} depend on the properties of the WIMP-radiation coupling only, through $\gamma(T)$, and on the cosmology through $H(T)$. As we show below, these expressions can be generalized to the case in which the cosmology is not in the form of a power-law model.

\section{Standard cosmology}\label{Standard cosmology}

In the standard radiation-dominated cosmology, it is common to parametrize the total energy density $\rho(T)$ and the total entropy density $s(T)$ in the universe with the so-called energy and entropy degrees of freedom $g(T)$ and $g_s(T)$, respectively, defined so that 
\begin{align}
\rho(T) = \frac{\pi^2}{30} \, g(T) \, T^4 ,
\qquad
s(T) = \frac{4\pi^2}{90} \, g_s(T) \, T^3 .
\end{align}
The Friedman equation and entropy conservation then give the relations
\begin{align} \label{define_Hrad}
H(T) = H^{\rm rad}(T) = T^2 \sqrt{\frac{4\pi^3 G}{45} g(T)} ,
\qquad
a^3 T^3 g_s(T) ={\rm const},
\end{align}
where $H^{\rm rad}(T)$ is the Hubble rate in the radiation-dominated cosmology. The scale factor depends on temperature as 
%\begin{align}
%\frac{da}{a} = - \frac{dT}{T} \left( 1 + \frac{1}{3} \frac{T}{g_s} \frac{d g_s}{dT} \right) ,
%\end{align}
\begin{align}
\frac{da}{a} = - \frac{dT}{T} \left( 1 + \frac{1}{3}  \frac{d \ln g_s}{d\ln T} \right) .
\end{align}
It follows that
\begin{align}
G(T,T') = G(a,a') = \sqrt{\frac{45}{\pi^3 G}} \int_{T}^{T'}  \frac{\gamma(T'')}{T^{\prime\prime 3}\sqrt{g(T'')}} \, \left( 1 + \frac{1}{3}  \frac{d \ln g_s(T'')}{d\ln T''} \right) \, dT'' .
\end{align}
and the WIMP kinetic temperature is
%\begin{align}
%T_\chi = T + \left( T_{\chi i}-T_i \right) \,\frac{T^2 g_s^{2/3}}{T_i^2 g_{si}^{2/3}}\,e^{-G(T,T_i)} - T^2 g_s^{2/3} \int_{a_i}^a e^{-G(a,a')}\,\frac{d}{dT'}\left[\frac{1}{T' g_{s}^{\prime 2/3}}\right]\, dT'.
%\end{align}
\begin{align}
T_\chi = T + \left( T_{\chi i}-T_i \right) \,\frac{T^2 g_s(T)^{2/3}}{T_i^2 g_{s}(T_i)^{2/3}}\,e^{-G(T,T_i)} -  \int_{T}^{T_i} e^{-G(T,T')}\,\frac{T^2 g_s(T)^{2/3}}{T^{\prime 2} g_{s}(T')^{2/3}} \, \left( 1 + \frac{2}{3}  \frac{d \ln g_s(T')}{d\ln T'} \right) \, dT'.
\end{align}
We have presented these formulas as they may be useful when implementing these expressions in dark matter numerical codes like DarkSUSY~\cite{darksusy} and micrOmegas~\cite{micromega}.

\section{Power-law cosmological model}\label{Power-law models}

In this section we consider models in which the Hubble rate $H$, the scale factor $a$, and the momentum relaxation rate $\gamma$ have a power law dependence on the plasma temperature $T$. More specifically, we consider a dependence of the Hubble rate on temperature of the form
\begin{equation} \label{H_powerlaw}
H(T) = \Hi\,\left(\frac{T}{\Ti}\right)^\nu,
\end{equation}
where $\nu$ is a positive constant, and $\Ti$ and $\Hi$ are the temperature of the plasma and the expansion rate at the time at which we start considering the cosmological model. For the dependence of the scale factor on the temperature we write
\begin{equation} \label{T_a_powerlaw}
a^{\alpha} \, T   = {\rm const}.
\end{equation}
Equating Eqs.~(\ref{H_powerlaw}) and~(\ref{T_a_powerlaw}), we obtain the relation
\begin{equation} \label{H_a_powerlaw}
H(a) = \Hi\,\left(\frac{\ai}{a}\right)^{\nu\,\alpha},
\end{equation}
where $\ai$ is the scale factor at temperature $\Ti$. 
Notice that, in the radiation-dominated cosmology for which $\nu=2$ and $\alpha=1$, the temperature of the plasma drops as $T \propto a^{-1}$, while the WIMP temperature drops at a faster rate $T_\chi \propto a^{-2}$. For the momentum relaxation rate $\gamma(T)$ we assume a power-law function of the form
\begin{equation} \label{damping}
\gamma(T) = \gi \left( \frac{T}{\Ti} \right) ^{4+n},
\end{equation}
where $\gi=\gamma(\Ti)$. In some models, the exponent $n$ is related to the low relative velocity $v$ of the forward WIMPs scattering amplitude $\mathcal{M}^{\rm forward}$ off particles in the plasma,
\begin{equation} \label{amplitude}
|\mathcal{M}^{\rm forward}|^2 = {\rm const} \, v^n . %c\,\left(\frac{T}{M_\chi}\right)^n.
\end{equation}
Finally, in power-law models, Eq.~\eqref{def_f} is given by
\begin{align}
\Upsilon = \frac{\gamma}{H} = \Ui  \left( \frac{T}{\Ti} \right) ^{4+n-\nu} = \Ui  \left( \frac{\ai}{a} \right) ^{\alpha(4+n-\nu)},
\end{align}
where $\Ui=\gi/\Hi$.

\subsection{Kinetic temperature}

Using the definition in Eq.~\eqref{gen_def_s} in the power-law model, we find
\begin{equation} \label{def_s}
s(a) = \begin{cases}
\displaystyle
\frac{2\,\Ui}{\alpha(4+n-\nu)}\,\left(\frac{\ai}{a}\right)^{\alpha(4+n-\nu)},&\hbox{for $4+n \neq \nu$},\\[1ex]
\displaystyle
-2\Ui\,\ln\left(\frac{a}{\ai}\right),&\hbox{for $4+n = \nu$}.
\end{cases}
\end{equation}
In the first line, we have fixed the arbitrary constant in the definition of $s(a)$ so that $s(a)$ is a power law.
It is interesting to observe that for $4+n \neq \nu$
\begin{align} \label{define_s}
s = \frac{2}{\alpha(4+n-\nu)} \, \frac{\gamma}{H} .
\end{align}
Plugging Eqs.~(\ref{H_a_powerlaw}),~(\ref{T_a_powerlaw}), and~(\ref{eq_G_powerlaw}) into Eq.~(\ref{sol3}),  computing the integrals,  using the identity
\begin{equation} \label{identity_gamma}
\Gamma(1+r,x) = r\,\Gamma(r,x) + x^r\,e^{-x},
\end{equation}
and defining
\begin{align} \label{define_lambda}
\lambda = \frac{2-\alpha}{\alpha\,(4+n-\nu)},
\end{align}
we find
%\begin{equation} \label{Tchi_powerlaw}
%T_\chi = \begin{cases}
%\Ti\,\left(\frac{s}{\si}\right)^{\frac{2}{\alpha(4+n-\nu)}}\,e^{s-\si}+T\,s^{\frac{2-\alpha}{\alpha(4+n-\nu)}}\,e^s\,\left[\Gamma\left(1-\frac{2-\alpha}{\alpha(4 + n - \nu)}, s\right) - \Gamma\left(1-\frac{2 -\alpha}{\alpha(4 + n - \nu)}, \si\right)\right],\quad\hbox{for $4+n \neq \nu$},\\
%\Ti\,\left(\frac{\ai}{a}\right)^{2+2\Ui} + \frac{2\Ui\,T}{2+2\Ui-\alpha}\,\left[1-\left(\frac{\ai}{a}\right)^{2+2\Ui-\alpha}\right],\quad\hbox{for $4+n = \nu$}.
%\end{cases}
%\end{equation}
\begin{align}
T_\chi & = T\,s^{\lambda}\,e^s\,\big[\Gamma\left(1-\lambda, s\right) + \lambda\,\Gamma\left(-\lambda, \si\right)\big],&&\hbox{for $4+n \neq \nu$},
 \label{Tchi_powerlaw_1}
 \intertext{and}
T_\chi & = \Ti\,\left(\frac{\ai}{a}\right)^{2+2\Ui} + \frac{2\Ui\,T}{2+2\Ui-\alpha}\,\left[1-\left(\frac{\ai}{a}\right)^{2+2\Ui-\alpha}\right],&&\hbox{for $4+n = \nu$}.
 \label{Tchi_powerlaw_2}
\end{align}
To the best of our knowledge, the expressions in Eqs.~(\ref{Tchi_powerlaw_1}-\ref{Tchi_powerlaw_2}) have never been derived for the case of an arbitrary power-law model. In Appendix~\ref{appendix2}, we show how to obtain the result in Eqs.~(\ref{Tchi_powerlaw_1}-\ref{Tchi_powerlaw_2}) from solving the differential Eq.~(\ref{mastereq1}) directly. 
%Thanks to the identity
%\begin{equation} \label{identity_gamma}
%\Gamma(1+r,x) = r\,\Gamma(r,x) + x^r\,e^{-x},
%\end{equation}
%we rewrite the first line of Eq.~(\ref{Tchi_powerlaw}) as
%\begin{equation} \label{Tchi_powerlaw1}
%T_\chi = T\,s^{\lambda}\,e^s\,\big[\Gamma\left(1-\lambda, s\right) + \lambda\,\Gamma\left(-\lambda, \si\right)\big],
%\end{equation}
%where
%\begin{align}
%\lambda = \frac{2-\alpha}{\alpha(4+n-\nu)}.
%\end{align}

If the initial scale factor $\ai$ is taken so far back in time that the WIMPs are initially tightly coupled to the primordial plasma, then $\gi \gg \Hi$ and $\si \to +\infty$, and we obtain 
\begin{equation} \label{Tchi_powerlaw2}
T_\chi = T\,s^{\lambda}\,e^s\,\Gamma\left(1-\lambda, s\right).
\end{equation}
Eq.~(\ref{Tchi_powerlaw2}) is a generalization of the relation obtained in Ref.~\cite{bertschinger} for any cosmological power-law model and for any value of the partial wave number $n$.

\subsection{$\Tkd$ in the power-law model}

Using the expressions in Eqs.~(\ref{H_powerlaw}) and~(\ref{damping}) and the definition of the temperature of the plasma at which the WIMP decouples kinetically, Eq.~\eqref{relation}, we obtain for $4+n\neq\nu$,
\begin{equation}\label{Tkd_bg}
\Tkd = \Ti  \, \Ui^{-\frac{1}{4+n-\nu}}
\end{equation}
while in the case $4+n = \nu$ we find the constraint $\gi = \Hi$, for which no solution for $\Tkd$ exists if this condition is not satisfied at all temperatures. In the literature, different definitions of the temperature of kinetic decoupling can be found. If we were to use the definition in Eq.~\eqref{relation_bertschinger}, we would obtain
\begin{equation} \label{Tkd_bertschinger}
\Tkd^B = 2^{1/(4+n-\nu)}\,\Tkd.
\end{equation}
Eq.~\eqref{define_s_R} below shows that the definition in Eq.~\eqref{Tkd_bertschinger} is obtained by setting the variable $s = 1$ in the standard cosmology, for the case of a $p$-wave.

Finally, the definition in Eq.~\eqref{Tkd_bringmann} applied to the power-law model discussed in this section gives $\Tkdb = 0$, unless in the particular case $\alpha = 1$ to which radiation-dominated cosmology belongs to. For this reason, we suggest to modify the equivalent definition in Eq.~(\ref{Tkd_bringmann}) as
\begin{equation} \label{Tkd_visinelli}
\Tkdv = \frac{\Ti^2}{T_\chi}\,\left(\frac{T}{\Ti}\right)^{\frac{2}{\alpha}}\bigg|_{T\to 0},
\end{equation}
Plugging Eq.~(\ref{Tchi_powerlaw2}) for $T_\chi$, valid for $4+n\neq \nu$, in the definition in Eq.~(\ref{Tkd_visinelli}), gives
\begin{equation} \label{result_Tkdv}
\Tkdv = \frac{s_i^{-\lambda}\,e^{-s}}{\Gamma\left(1-\lambda, s\right)}\bigg|_{T\to 0}\,\Ti = \frac{s_i^{-\lambda}}{\Gamma\left(1-\lambda\right)}\,\Ti,
\end{equation}
where $s_i$ is a temperature-independent quantity defined by the relation in Eq.~\eqref{define_si} as
\begin{align} \label{define_si}
s_i = \frac{2}{\alpha(4+n-\nu)} \,\Upsilon_i .
\end{align}
As anticipated in Sec.~\ref{Temperature of kinetic decoupling}, the three results in Eqs.~\eqref{Tkd_bg},~\eqref{Tkd_bertschinger}, and~\eqref{result_Tkdv} have the same dependence on $\Ti$, up to a numerical constant. However, the definition in Eq.~(\ref{Tkd_visinelli}) depends in principle on the moment at which the temperature of KD is computed, unlike the definition we adopted in Eq.~(\ref{relation}) which only depends on the properties of the WIMP-radiation coupling through $\gamma(T)$ and on the cosmology through $H(T)$, and can be generalized to the case in which the cosmology is not in the form of a power-law model.

\subsection{Radiation-dominated cosmology}

In the standard radiation-dominated model $\alpha = 1$ and $\nu = 2$, Eq.~(\ref{define_lambda}) gives $\lambda = 1/(2+n)$. Using the identity in Eq.~(\ref{identity_gamma}), the limit for $T_\chi$ expressed in Eq.~(\ref{Tchi_powerlaw2}) gives the same result as Eq.~(4) in Ref.~\cite{bringmann},
\begin{equation} \label{Tchi_powerlaw_bringmann}
T_\chi = T\,s_R^{\frac{1}{2+n}}\,e^{s_R}\,\Gamma\left(\frac{1+n}{2 + n}, s_R\right) = T\,\left[1 - \frac{1}{2+n}\,e^{s_R}\,s_R^{\frac{1}{2+n}}\,\Gamma\left(-\frac{1}{2+n},s_R\right)\right],
\end{equation}
where the variable $s$ defined in general through Eq.~(\ref{gen_def_s}) reduces in the radiation-dominated cosmology to
\begin{align} \label{define_s_R}
s_R = \frac{2}{2+n}\,\frac{\gamma}{H}.
\end{align}
In addition, when we consider $p$-wave scattering ($n = 2$) in a radiation-dominated model, then $s_R = \gamma/2H$, and we obtain Eq.~(12) in Ref.~\cite{bertschinger},
\begin{equation} \label{Tchi_powerlaw_bertschinger}
T_\chi = T\,s_R^{1/4}\,e^{s_R}\,\Gamma\left(\frac{3}{4}, s_R\right).
\end{equation}

%{\it 
%Note that with the definition $\gamma(\Tkd)=H(\Tkd)$ in the last model, one has $\Tkd = \Ti/\Ui^{1/4}$ and 
%\begin{align}
%T_\chi = \frac{\Gamma\left(\frac{3}{4}\right)}{2^{1/4}} \, \frac{T^2}{\Tkd} 
%\end{align}
%at small $T$. In a power-law model,  $\gamma(\Tkd)=H(\Tkd)$ implies $\Tkd = \Ti/\Ui^{1/(4+n-\nu)}$ and
%\begin{align}
%T_\chi =  \left( \frac{2\lambda}{2-\alpha} \right)^\lambda \, \Gamma(1-\lambda) \, \frac{T^{2/\alpha}}{\Tkd^{(2/\alpha)-1}} .
%\end{align}
%Notice
%\begin{align}
%a^2 \, T_\chi = \left( \frac{2\lambda}{2-\alpha} \right)^\lambda \, \Gamma(1-\lambda) \,  a_{\rm kd}^2\, \Tkd.
%\end{align}
%}
%
\subsection{Late time behavior}

When the plasma temperature is much smaller than $T_i$, the late-time behavior of Eq.~(\ref{Tchi_powerlaw_1}) gives
\begin{equation} \label{Tchi_powerlaw2_noC}
T_\chi = \Ti\,s_i^\lambda\,\left(\frac{T}{\Ti}\right)^{\frac{2}{\alpha}}\,\Gamma\left(1-\lambda\right).
\end{equation}
In a cosmological model that approaches the radiation-dominated scenario where $\alpha = 1$ and $\nu = 2$, Eq.~\eqref{Tchi_powerlaw2_noC} reads
\begin{equation} \label{Tchi_stdcosm_noC}
T_\chi = \frac{T^2}{\Ti}\,\left(\frac{2\,\Ui}{2+n}\right)^{\frac{1}{2+n}}\,\Gamma\left(\frac{1+n}{2 + n}\right).
\end{equation}
We compare this result with the theoretical behavior~\cite{bringmann}
\begin{equation}\label{def_Bringmann}
T_\chi^{\rm th} = \frac{T^2}{\Tstd}\,\left(\frac{2}{2+n}\right)^{\frac{1}{2+n}}\,\Gamma\left(\frac{1+n}{2+n}\right),
\end{equation}
where $\Tstd$ is the temperature of kinetic decoupling in the radiation-dominated cosmology,
\begin{equation} \label{Tkd_std}
\Tstd = T_i\,\left(\frac{H^{\rm rad}(T_i)}{\gamma_i}\right)^{\frac{1}{2+n}},
\end{equation}
and $H^{\rm rad}(T)$ has been defined in Eq.~\eqref{define_Hrad}. This latter equation can be stated in terms of the function $\Ui$ in Eq.~(\ref{Tkd_bg}) as
\begin{equation} \label{TRH_Tkd}
\Ti = \Tstd\,\Ui^{\frac{1}{2+n}}.
\end{equation}
This relation is also obtained by comparing the result in Eq.~\ref{Tchi_stdcosm_noC} with the theoretical Eq.~\eqref{def_Bringmann}. We rewrite Eq.~\eqref{TRH_Tkd} in terms of the temperature of kinetic decoupling $\Tkd$ by using the relation in Eq.~\eqref{Tkd_bg} in the form
\begin{equation}\label{Tkd_bg}
\Ui = \left(\frac{T_i}{\Tkd}\right)^{4+n-\nu},
\end{equation}
as
\begin{equation} \label{TRH_Tkd_noC}
\Tkd = \left(\frac{\Tstd^{n+2}}{\Ti^{\nu-2}}\right)^{\frac{1}{4+n-\nu}} = T_i\,\left(\frac{H^{\rm rad}(T_i)}{\gamma_i} \right)^{\frac{1}{4+n-\nu}}.
\end{equation}
Eq.~\eqref{TRH_Tkd_noC} gives the temperature of the WIMP kinetic decoupling in a generic cosmological model, which might differ from the radiation-dominated scenario at the time of decoupling. Notice that, in the particular case in which the decoupling occurs in a radiation-dominated scenario ($\nu = 2$), Eq.~\eqref{TRH_Tkd_noC} gives
\begin{equation} \label{TRH_Tkd_std}
\Tkd = \Tstd.
\end{equation}
In the following, we discuss the decoupling of WIMPs in a broken power law cosmological model, where a generic pre-BBN cosmology takes place before $\Ti$, after which standard radiation-dominated cosmology begins.

\section{Broken-power-law cosmological model}\label{Broken power-law model}
\subsection{Generic pre-BBN cosmology}

The relic density and velocity distribution of WIMPs depend on the characteristics of the Universe before Big Bang Nucleosynthesis (BBN), which occurred at temperature above $T_{\rm BBN} \sim 1{\rm~MeV}$. Since this is an epoch from which we have no data, it is possible that the expansion rate of the Universe prior BBN differed from that of the radiation-dominated period, with the transit between the two cosmological models occurring at a reheating temperature $\TRH$ which must lie above~\cite{kawasaki1999, kawasaki2000, hannestad2004, ichicawa2005, debernardis2008}
\begin{equation} \label{TRH_limit}
\TRH > 4\,{\rm MeV}.
\end{equation}
For this reason, in this section we generalize the power-law model discussed in the previous section by considering a generic scenario in which the Universe is dominated by some form of energy before cooling down to a reheating temperature $\TRH$, after which standard radiation-dominated cosmology takes place. For this, we assume a dependence of the Hubble rate on temperature in the form
\begin{equation} \label{H_nonstandard}
H(T) = \HRH\,
\begin{cases}
\left(\frac{T}{\TRH}\right)^\nu, & \hbox{for $T > \TRH$} ,\\
\left(\frac{T}{\TRH}\right)^2, & \hbox{for $T < \TRH$},
\end{cases}
\end{equation}
where $\nu$ is a positive constant, $\TRH$ is the reheating temperature, and $\HRH = H(\TRH)$. In this scenario, the dependence of temperature on the scale factor is a power law of the form
\begin{equation} \label{T_a}
T(a) = \TRH\,
\begin{cases}
\left(\frac{\aRH}{a}\right)^\alpha, & \hbox{for $a < \aRH$} ,\\
\frac{\aRH}{a}, & \hbox{for $a > \aRH$},
\end{cases}
\end{equation}
where $\aRH$ is the scale factor when $T = \TRH$. In the radiation-dominated cosmology the temperature of the plasma drops as $T \propto a^{-1}$, while the WIMP temperature drops at a faster rate $T_\chi \propto a^{-2}$. Equating Eqs.~(\ref{H_nonstandard}) and~(\ref{T_a}), we obtain the relation for the Hubble rate in terms of the scale factor,
\begin{equation} \label{H_a}
H(a) = \HRH \begin{cases}
\left(\frac{\aRH}{a}\right)^{\nu\,\alpha}, & \hbox{for $a < \aRH$} ,\\
\left(\frac{\aRH}{a}\right)^2, & \hbox{for $a > \aRH$} ,
\end{cases}
\end{equation}
Incidentally, a relation between $\fRH \equiv \gamma(\TRH)/H(\TRH)$ and $\Tkd$ is obtained using the definition for the temperature of kinetic decoupling introduced in Eq.~(\ref{relation}), in the form $\Upsilon(\Tkd) = 1$, and with Eqs.~(\ref{damping}) and~(\ref{H_nonstandard}),
\begin{equation} \label{def_gammaRH}
\fRH = 
\begin{cases}
\left(\frac{\TRH}{\Tkd}\right)^{4+n-\nu},& \hbox{for $\Tkd > \TRH$},\\
\left(\frac{\TRH}{\Tkd}\right)^{2+n},& \hbox{for $\Tkd < \TRH$}.
\end{cases}
\end{equation}
Plugging Eqs.~(\ref{T_a}-\ref{H_a}) into Eq.~(\ref{behavior_asymptotic}) we find, for $a < \aRH$,
\begin{align} \label{Tkin_befRH}
T_\chi = \begin{cases}
T\,s^{\lambda}\,e^s\,\Gamma\left(1-\lambda, s\right), & \hbox{for $4+n \neq \nu$},\\
\frac{2\fRH\,T}{2+2\fRH-\alpha}, & \hbox{for $4+n = \nu$},
\end{cases}
\end{align}
where $s$ has been defined in Eq.~(\ref{define_s}) and $\lambda$ in Eq.~(\ref{define_lambda}).
% and we have introduced, for $4+n\neq\nu$,
%\begin{equation} \label{def_s_broken}
%s=s(a) = \hs\,\left(\frac{\aRH}{a}\right)^{\alpha\,(4+n-\nu)}, \quad\hbox{with}\quad \hs = \frac{2}{\alpha\,(4+n-\nu)}\,\fRH.
%\end{equation}
After reheating $a \geq \aRH$, we obtain
\begin{equation} \label{Tkin_aftRH}
T_\chi = \frac{T^2}{\TRH}\,e^{s_R}\,\hsR^{\frac{1}{2+n}}\,\left[\Gamma\left(\frac{1+n}{2+n},s_R \right) + C_n\right],
\end{equation}
where $s_R$ has been defined in Eq.~(\ref{define_s_R}), we set
\begin{equation} \label{def_u_broken}
\hsR = \frac{2}{2+n}\,\fRH, \quad \hs = \frac{2}{\alpha\,(4+n-\nu)}\,\fRH,
\end{equation}
and where the constant matching the two cosmologies at $a = \aRH$ is
\begin{align}
C_n = \hs^{\lambda}\,\hsR^{\frac{-1}{2+n}}\,e^{\hs-\hsR}\,\Gamma\left(1 - \lambda, \hs\right) - \Gamma\left(\frac{1+n}{2+n}, \hsR \right),\quad\hbox{for $4+n \neq \nu$},
\label{const_Tkin_1}
 \intertext{and}
C_n = \hsR^{\frac{-1}{2+n}}\,e^{-\hsR}\,\frac{2\fRH}{2+2\fRH-\alpha} - \Gamma\left(\frac{1+n}{2+n}, \hsR \right),\quad\hbox{for $4+n = \nu$}.
 \label{const_Tkin_2}
\end{align}
For high temperature $T_\chi \approx T$, Eq.~(\ref{Tkin_befRH}) shows the limit discussed in Eq.~(\ref{Tchi_powerlaw2}). When the plasma temperature approaches $T\to 0$, Eq.~(\ref{Tkin_aftRH}) gives
\begin{equation} \label{limit_Ttozero}
T_\chi = \frac{T^2}{\TRH}\,\hsR^{\frac{1}{2+n}}\,\left[\Gamma\left(\frac{1+n}{2+n}\right) + C_n\right].
\end{equation}
Comparing the result in Eq.~(\ref{limit_Ttozero}) with the theoretical expression in Eq.~(\ref{def_Bringmann}) gives
\begin{equation} \label{TRH_Tkd_broken}
\TRH = \Tstd\,\fRH^{\frac{1}{2+n}}\,\left[1 + \frac{C_n}{\Gamma\left(\frac{1+n}{2+n}\right)}\right].
\end{equation}
This expression is similar to Eq.~(\ref{TRH_Tkd}), with the additional inclusion of the matching constant $C_n$ which comes from the extra pre-BBN cosmology.

\subsection{Low temperature reheating cosmology} \label{LTR}

\begin{figure}[h!]
  \includegraphics[width=12cm]{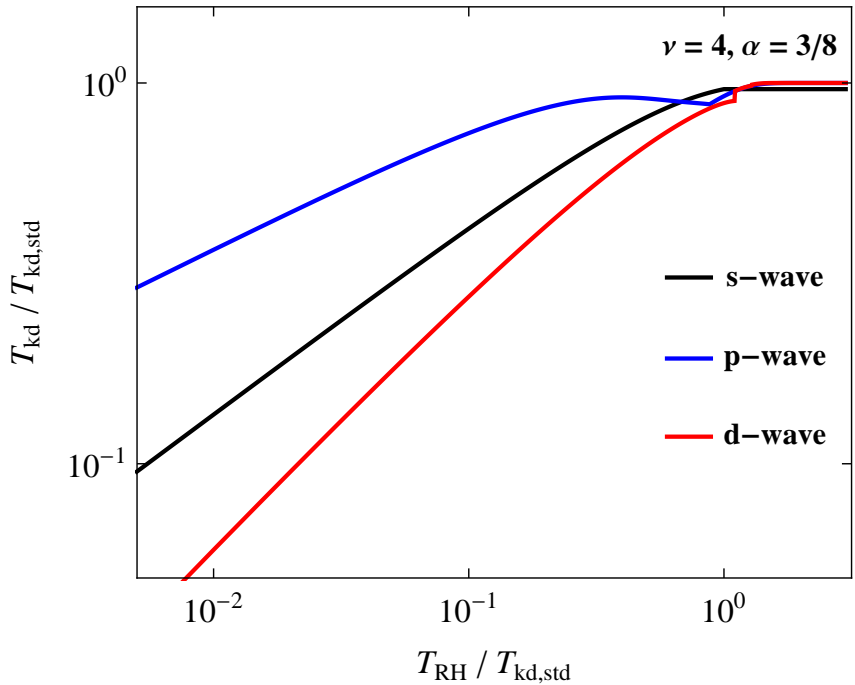}
\caption{The temperature of kinetic decoupling $\Tkd$ as a function of the reheating temperature $\TRH$ in the LTR cosmology. Both temperatures are given in units of $\Tstd$. We show results from Eqs.~(\ref{TRH_Tkd_LTR_1}) and~\eqref{TRH_Tkd_LTR_2} for $s$-wave (black), $p$-wave (blue), and $d$-wave scattering (red).}
\label{figureTRH_LTR}
\end{figure}

\begin{figure}[h!]
\includegraphics[width=12cm]{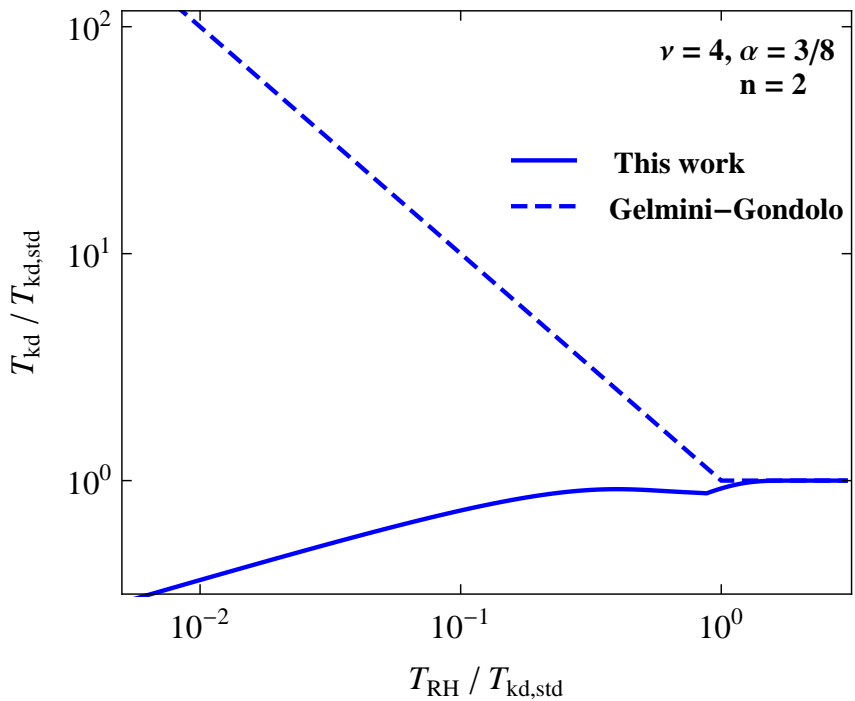}
\caption{The temperature of kinetic decoupling $\Tkd$ as a function of the reheating temperature $\TRH$, for a dominating $p$-wave mode in the pre-BBN LTR cosmology. Both temperatures are given in units of $\Tstd$. Results are shown both using the Gelmini-Gondolo model in Eq.~(\ref{Tkd_GG}) (blue dashed line), and the results derived in this paper from Eq.~(\ref{TRH_Tkd_LTR_n2}) (blue solid line).}
\label{figure:Tkd_comparison}
\end{figure}

In the Low Temperature Reheating (LTR) scenario \cite{turner_LTR, scherrer1985, dine, steinhardt}, the energy density of a massive scalar field $\phi$ dominates the evolution of the Universe in the pre-Big Bang Nucleosynthesis (BBN) epoch, after which standard radiation-dominated cosmology begins. The evolution of the Universe during the LTR stage is non-adiabatic: in fact, in the LTR model, the Universe is constantly reheated by the decay of the massive particle $\phi$. This scenario is modeled by the Hubble rate in Eq.~(\ref{H_nonstandard}), with $\nu = 4$, and by the scale factor in Eq.~(\ref{H_a}), with $\alpha = 3/8$. In the LTR scenario, Eq.~(\ref{TRH_Tkd_broken}) for the reheating temperature as a function of $\fRH$, with the constants in Eqs.~\eqref{const_Tkin_1} and~\eqref{const_Tkin_2}, reads
\begin{align}
\frac{\TRH}{\Tstd} = \fRH^{\frac{1}{2+n}}\,\left\{1+  \frac{1}{\Gamma\left(\frac{1+n}{2+n}\right)}\,\left[\frac{\hs^{\frac{13}{3n}}\,e^{\hs}}{\hsR^{\frac{1}{2+n}}\,e^{\hsR}}\,\Gamma\left(\frac{3n-13}{3n}, \hs\right) - \Gamma\left(\frac{1+n}{2+n}, \hsR \right)\right]\right\},\quad\hbox{for $n \neq 0$},
\label{TRH_Tkd_LTR_1}
 \intertext{and}
\frac{\TRH}{\Tstd} = \fRH^{1/2}\,\,\left\{1+  \frac{1}{\Gamma\left(\frac{1}{2}\right)}\,\left[\frac{1}{\hsR^{\frac{1}{2}}\,e^{\hsR}}\,\frac{2\fRH}{2+2\fRH-\alpha} - \Gamma\left(\frac{1}{2}, \hsR \right)\right]\right\},\quad\hbox{for $n=0$}.
\label{TRH_Tkd_LTR_2}
\end{align}
Fig.~\ref{figureTRH_LTR} shows the value of the temperature of kinetic decoupling $\Tkd$ as a function the reheating temperature $\TRH$ in the LTR cosmology, obtained from solving Eqs.~\eqref{TRH_Tkd_LTR_1} and~\eqref{TRH_Tkd_LTR_2}. Both temperatures are given in units of $\Tstd$. We considered the case of a dominating $s$-wave (blue), $p$-wave (blue), or $d$-wave scattering (red).
Gelmini and Gondolo \cite{gondolo_gelmini} recovered a relation between $\Tkd$ and $\TRH$ in the case of the LTR cosmology and for $p$-wave scattering (see their Eq.~(6) in Ref.~\cite{gondolo_gelmini})
\begin{equation}
\Tkd^{\rm GG} =
\begin{cases}
\frac{\Tstd^2}{\TRH}, & \hbox{for $\Tstd > \TRH$},\\
\Tstd,& \hbox{for $\Tstd < \TRH$}.
\end{cases}
\label{Tkd_GG}
\end{equation}
where $\Tstd$ has been defined in Eq.~(\ref{Tkd_std}). Within the framework of the present paper, using Eq.~\eqref{TRH_Tkd_LTR_1} with $n = 2$ gives
\begin{equation}
\Tkd =
\begin{cases}
\frac{\Tstd^2}{\TRH}\,\left[1+  \frac{1}{\Gamma\left(\frac{3}{4}\right)}\,C_2\right]^2,& \hbox{for $\Tkd > \TRH$},\\
\Tstd\,\left[1+  \frac{1}{\Gamma\left(\frac{3}{4}\right)}\,C_2\right],& \hbox{for $\Tkd < \TRH$}.
\end{cases}
\label{TRH_Tkd_LTR_n2}
\end{equation}
The main difference between Eqs.~\eqref{Tkd_GG} and~\eqref{TRH_Tkd_LTR_n2} stands in the appearance of the constant $C_2$. In Fig.~\ref{figure:Tkd_comparison}, we show the relation between $\Tkd$ and $\TRH$ for a dominating $p$-wave WIMP particle, using the Gelmini-Gondolo model in Eq.~(\ref{Tkd_GG}) (blue dashed line), and our result in Eq.~(\ref{TRH_Tkd_LTR_n2}) (blue solid line).

\subsection{Kination cosmology} \label{kination}

In the kination cosmology scenario \cite{ford}, the expansion of the Universe before the standard radiation-dominated cosmology begins is driven by the kinetic energy of a scalar field. We model the kination scenario by setting the values for the pre-BBN cosmology $\alpha = 1$ and $\nu = 3$ in Eqs.~(\ref{H_nonstandard}) and~(\ref{H_a}). With these values, the condition $4+n \neq \nu$ is always satisfied for all $n>0$, and the reheating temperature in Eq.~\eqref{TRH_Tkd_broken} in the kination cosmology is
\begin{equation} \label{TRH_Tkd_kin}
\TRH = \Tstd\,\fRH^{\frac{1}{2+n}}\,\left\{1 + \frac{1}{\Gamma\left(\frac{1+n}{2+n}\right)}\,\left[\frac{\hs^{\frac{1}{1+n}}\,e^{\hs}}{\hsR^{\frac{1}{2+n}}\,e^{\hsR}}\,\Gamma\left(\frac{n}{1+n}, \hs\right) - \Gamma\left(\frac{1+n}{2+n}, \hsR \right)\right]\right\}.
\end{equation}

\begin{figure}[h!]
\includegraphics[width=12cm]{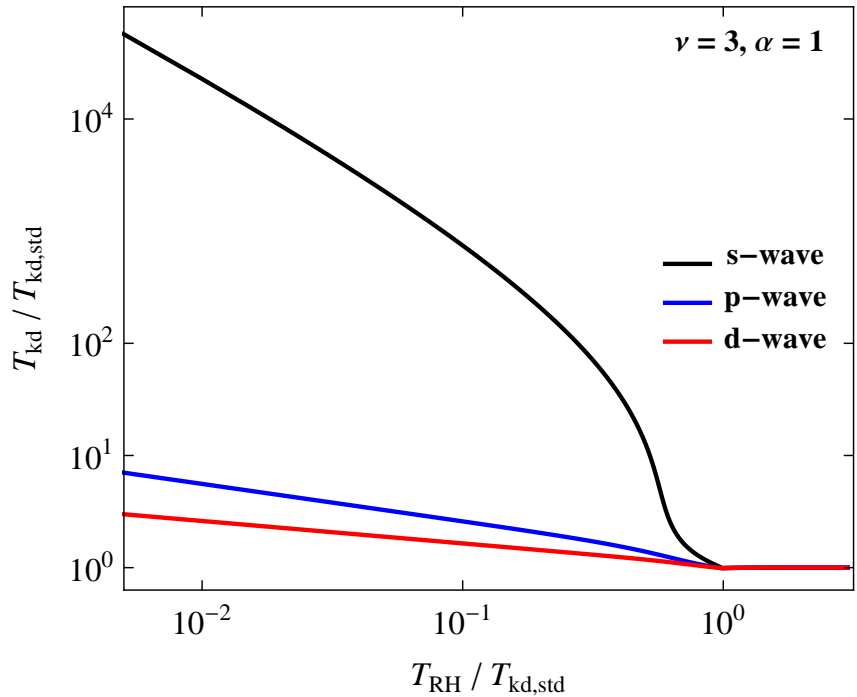}
\caption{The temperature of kinetic decoupling $\Tkd$ as a function of the reheating temperature $\TRH$ in the kination cosmology. Both temperatures are given in units of $\Tstd$. We show results from Eq.~(\ref{TRH_Tkd_kin}) for $s$-wave (black), $p$-wave (blue), and $d$-wave scattering (red).}
\label{figureTkd_kin}
\end{figure}

Fig.~\ref{figureTkd_kin} shows the value of the temperature of kinetic decoupling $\Tkd$ as a function the reheating temperature $\TRH$ in the kination cosmology, obtained from solving Eq.~\eqref{TRH_Tkd_kin}. Both temperatures are given in units of $\Tstd$. We considered the case of a dominating $s$-wave (black), $p$-wave (blue), or $d$-wave scattering (red).

\section{Summary} \label{Summary}

In Eq.~(\ref{sol2_2}), we presented a general expression that gives the value of the WIMP kinetic temperature $T_\chi$ in terms of the temperature of the Universe $T$. This generic result has been specialized to the case of the standard cosmology in Sec.~\ref{Standard cosmology}, in view of possible future applications to numerical models. In addition, we have presented the expression for $T_\chi$ in the case of a power-law cosmology in Sec.~\ref{Power-law models}, and for a broken power-law model in Sec.~\ref{Broken power-law model}. In Sec.~\ref{LTR}, we have discussed the numerical results in the case of the LTR cosmology, and we have compared our results with those obtained in Ref.~\cite{gondolo_gelmini} results, finding a discrepancy between our values of $\Tkd$ as a function of $\TRH$ for low values of $\TRH$. We have imputed this discrepancy to the appearance of the constant $C_n$ in Eq.~(\ref{TRH_Tkd_broken}), which come from the matching conditions between the LTR and the standard cosmologies. Results for the kination cosmology have been discussed in see Sec.~\ref{kination}. 

\begin{acknowledgements}
P.G. was partially supported by NSF award PHY-1068111 and PHY-1415974.
\end{acknowledgements}

\appendix

\section{Direct solution of Eq.~(\ref{mastereq1})} \label{appendix2}

In the scenario described in Sec.~\ref{Power-law models}, we solve the differential Eq.~(\ref{mastereq1}) by first writing $T_\chi$ as a function of the plasma temperature $T$,
\begin{equation} \label{mastereq_a}
\frac{dT}{dt}\,\frac{dT_\chi}{dT} + 2\,\left(H+\gamma(T)\right)\, T_\chi = 2\gamma(T)\,T.
\end{equation}
Here, $dT/dt$ is obtained by taking the time derivative of $T$ in Eq.~(\ref{T_a_powerlaw}) and using Eq.~(\ref{H_powerlaw}),
\begin{equation} \label{dT_dt}
\frac{dT}{dt} = -\alpha\,H\,T = -\alpha\,H_i\,T_i\,\left(\frac{T}{T_i}\right)^{\nu+1}.
\end{equation}
Using the result in Eq.~(\ref{dT_dt}) and Eq.~(\ref{damping}), Eq.~(\ref{mastereq_a}) reads
\begin{equation} \label{mastereq_appendix}
\frac{dT_\chi}{dT} - \frac{2}{\alpha\,T}\,\left[1 + \Ui\,\left(\frac{T}{T_i}\right)^{4+n-\nu}\,\right] \,T_\chi + \frac{2\,\Ui}{\alpha}\,\left(\frac{T}{T_i}\right)^{4+n-\nu} = 0.
\end{equation}
When $4+n \neq \nu$, we switch to the variable $s$ defined in Eq.~(\ref{def_s}), expressed in terms of the temperature instead of the scale factor,
\begin{equation} \label{s_a}
s = s(T) = \frac{2\,\Ui}{\alpha\,(4+n-\nu)}\,\left(\frac{T}{T_i}\right)^{4+n-\nu}.
\end{equation}
Rearranging Eq.~\eqref{mastereq_appendix}, we obtain
\begin{equation} \label{mastereq_appendix1}
\frac{dT_\chi}{ds} - \left[\frac{2}{\alpha\,(4+n-\nu)\,s} + 1\right] \,T_\chi + T_i\,\left[\frac{\alpha\,(4+n-\nu)\,s}{2\,\Ui}\right]^{\frac{1}{4+n-\nu}} = 0,
\end{equation}
whose general solution is
\begin{equation} \label{solution}
T_\chi = T\,\left(\frac{s}{s_i}\right)^\lambda\,e^{s-s_i}+T\,s^{\lambda}\,e^s\,\left[\Gamma\left(1-\lambda, s\right) - \Gamma\left(1-\lambda, s_i\right)\right],
\end{equation}
Eq.~\eqref{solution} is identical to Eq.~\eqref{Tchi_powerlaw_1} once the identity in Eq.~\eqref{identity_gamma} is used.

When $4+n = \nu$, Eq.~(\ref{mastereq_appendix}) reads
\begin{equation} \label{mastereq_appendix_zero}
\frac{dT_\chi}{dT} - \frac{2}{\alpha\,T} \,\left(1 + \Ui\right) \,T_\chi + \frac{2\Ui}{\alpha} = 0,
\end{equation}
and the solution is given by Eq.~\eqref{Tchi_powerlaw_2},
\begin{equation} \label{solution_zero}
T_\chi = T_i\,\left(\frac{a_i}{a}\right)^{2+2\Ui} + \frac{2\,\Ui\,T}{2+2\Ui-\alpha}\,\left[1-\left(\frac{a_i}{a}\right)^{2+2\Ui-\alpha}\right].
\end{equation}
To sum up, the solutions expressed in Eqs.~(\ref{solution}) and~(\ref{solution_zero}), obtained by solving the differential Eq.~(\ref{mastereq_a}), are equivalent to Eq.~(\ref{Tchi_powerlaw_1}) and~(\ref{Tchi_powerlaw_2}) respectively, obtained by solving the integral Eq.~(\ref{sol2}).


\begin{thebibliography}{50}

\bibitem{komatsu} E.~Komatsu {\it et al.} [WMAP Collaboration], Astrophys.\ J.\ Suppl.\ {\bf 180}, 330 (2009).

\bibitem{planck}  P.~A.~R.~Ade {\it et al.}  [Planck Collaboration], arXiv:1303.5076.

\bibitem{zwicky} F.~Zwicky, Phys.\ Acta {\bf 6}, 110 (1933).

\bibitem{kolb_book} E.~W.~Kolb and M.~S.~Turner, Addison-Wesley (1990).

\bibitem{jungman1996} G.~Jungman, M.~Kamionkowski, and K.~Griest, Phys.\ Rept.\ {\bf 267}, 195 (1996) [\href{http://arxiv.org/abs/hep-ph/9506380}{hep-ph/9506.380}].

\bibitem{bertone2005} G.~Bertone, D.~Hooper, and J.~Silk, Phys.\ Rept.\ {\bf 405}, 279 (2005) [\href{http://arxiv.org/abs/hep-ph/0404175}{hep-ph/0404.175}].

\bibitem{kuhlen2012} M.~Kuhlen, M.~Vogelsberger, and R.~Angulo, Phys.\ Dark Univ.\ {\bf 1}, 50 (2012) [\href{http://arxiv.org/abs/1209.5745}{astro-ph/1209.5745}].

\bibitem{hofmann} S.~Hofmann, D.~J.~Schwarz, and H.~Stoecker, Phys.\ Rev.\ D {\bf 64}, 083507 (2001)  [\href{http://arxiv.org/abs/astro-ph/0104173}{astro-ph/0104173}].

\bibitem{chen} X.~Chen, M.~Kamionkowski, and X.~Zhang, Phys.\ Rev.\ D {\bf 64}, 021302 (2001) [\href{http://arxiv.org/abs/astro-ph/0103452}{astro-ph/0103452}].

\bibitem{berezinsky} V.~Berezinsky, V.~Dokuchaev, and Y.~Eroshenko, Phys.\ Rev.\ D {\bf 68}, 103003 (2003) [\href{http://arxiv.org/abs/astro-ph/0301551}{astro-ph/0301551}].

\bibitem{green} A.~M.~Green, S.~Hofmann, and D.~J.~Schwarz, Mon.\ Not.\ Roy.\ Astron.\ Soc.\ {\bf 353}, L23 (2004); JCAP {\bf 0508}, 003 (2005) [\href{http://arxiv.org/abs/astro-ph/0503387}{astro-ph/0503387}].

\bibitem{bertschinger} E.~Bertschinger, Phys.\ Rev.\ D {\bf 74}, 063509 (2006) [\href{http://arxiv.org/abs/astro-ph/0607319}{astro-ph/0607319}].

\bibitem{bringmann} T.~Bringmann and S.~Hofmann, JCAP {\bf 0407}, 016 (2007) [\href{http://arxiv.org/abs/hep-ph/0612238}{hep-ph/0612238}].

\bibitem{kasahara} J.~Kasahara, Ph.\ D.\ dissertation, University of Utah (2009; ISBN 9781109295320; UMI microform 3368246).

\bibitem{bi} X.~J.~Bi, P.~F.~Yin, and Q.~Yuan, Phys.\ Rev.\ D {\bf 85}, 043526 (2012) [\href{http://arxiv.org/abs/1106.6027}{hep-ph/1106.6027}].

\bibitem{gondolo} P.~Gondolo, J.~Hisano, K.~Kadota, \href{http://arxiv.org/abs/1205.1914}{arXiv:1205.1914} [hep-ph].


\bibitem{schmid} C.~Schmid, D.~J.~Schwarz, and P.~Widerin, Phys.\ Rev.\ D {\bf 59}, 043517 (1999) [\href{http://arxiv.org/abs/astro-ph/9807257}{astro-ph/9807257}].

\bibitem{boehm} C.~Boehm, P.~Fayet, and R.~Schaeffer, Phys.\ Lett.\ B {\bf 518}, 8 (2001) [\href{http://arxiv.org/abs/astro-ph/0012504}{astro-ph/0012504}].

\bibitem{loeb} A.~Loeb and M.~Zaldarriaga, Phys.\ Rev.\ D {\bf 71}, 103520 (2005) [\href{http://arxiv.org/abs/astro-ph/0504112}{astro-ph/0504112}].

\bibitem{profumo} S.~Profumo, K.~Sigurdson, and M.~Kamionkowski, Phys.\ Rev.\ Lett. {\bf 97}, 031301 (2006) [\href{http://arxiv.org/abs/astro-ph/0603373}{astro-ph/0603373}].

\bibitem{gondolo_gelmini} G.~B.~Gelmini and P.~Gondolo, JCAP {\bf 0810} 002, (2008) [\href{http://arxiv.org/abs/0803.2349}{astro-ph/0803.2349}].

\bibitem{aarssen} L.~G.~van~den~Aarssen, T.~Bringmann, and Y.~CGoedecke, Phys.\ Rev.\ D {\bf 85}, 123512 (2012) [\href{http://arxiv.org/abs/1202.5456}{hep-ph/1202.5456}].

\bibitem{cornell} J.~M.~Cornell and S.~Profumo, JCAP {\bf 1206}, 011 (2012) [\href{http://arxiv.org/abs/1203.1100}{hep-ph/1203.1100}].

\bibitem{darksusy} P.~Gondolo, J.~Edsj{\"o}, P.~Ullio, L.~Bergstr{\"o}m, M.~Schelke, and E.~A.~Baltz, JCAP {\bf 0407}, 008  (2004) [\href{http://arxiv.org/abs/astro-ph/0406204}{astro-ph/0406204}].

\bibitem{micromega} G.~Belanger, F.~Boudjema, A.~Pukhov, and A.~Semenov, Comput.\ Phys.\ Commun.\ {\bf 185}, 960 (2014) [\href{http://arxiv.org/abs/1305.0237}{hep-ph/1305.0237}].

\bibitem{kawasaki1999} M.~Kawasaki, K.~Kohri, and N.~Sugiyama, Phys.\ Rev.\ Lett.\ {\bf 82}, 4168 (1999) [\href{http://arxiv.org/abs/astro-ph/9811437}{astro-ph/9811.437}].

\bibitem{kawasaki2000} M.~Kawasaki, K.~Kohri, and N.~Sugiyama, Phys.\ Rev.\ D {\bf 62}, 023506 (2000) [\href{http://arxiv.org/abs/astro-ph/0002127}{astro-ph/0002.127}].

\bibitem{hannestad2004} S.~Hannestad, Phys.\ Rev.\ D {\bf 70}, 043506 (2004) [\href{http://arxiv.org/abs/astro-ph/0403291}{astro-ph/0403.291}].

\bibitem{ichicawa2005} K.~Ichikawa, M.~Kawasaki, and F.~Takahashi,  Phys.\ Rev.\ D {\bf 72}, 043522 (2005) [\href{http://arxiv.org/abs/astro-ph/0505395}{astro-ph/0505.395}].

\bibitem{debernardis2008} F.~De Bernardis, L.~Pagano, and A.~Melchiorri, Astrop.\ Phys.\ {\bf 30}, 192 (2008).

\bibitem{turner_LTR} M.~S.~Turner, Phys.\ Rev.\ D {\bf 28}, 1243 (1983)

\bibitem{scherrer1985} R.~J.~Scherrer, M.~S.~Turner, Phys.\ Rev.\ D {\bf 31}, 681 (1985).

\bibitem{dine} M.~Dine, W.~Fishler, Phys.\ Lett.\ B {\bf 120}, 137 (1983).

\bibitem{steinhardt} P.~J.~Steinhardt, M.~S.~Turner, Phys.\ Lett.\ B {\bf 129}, 51 (1983).

\bibitem{ford} L.~H.~Ford, Phys.\ Rev.\ D {\bf 35}, 2955 (1987).

\end{thebibliography}
\end{document}